\documentclass[reprint,amsmath,amssymb,nofootinbib,aps,pra]{revtex4-2}
\pdfoutput=1
\usepackage{graphicx}
\usepackage{dcolumn}
\usepackage{bm}
\usepackage[breaklinks=true,colorlinks=true,linkcolor=blue,urlcolor=blue,citecolor=blue]{hyperref}
\usepackage{multirow}
\usepackage[dvipsnames]{xcolor}
\usepackage[pscoord]{eso-pic}
\usepackage[normalem]{ulem}
\usepackage{slashed}
\usepackage{makecell}

\newcommand{\placetextbox}[3]{
	\setbox0=\hbox{#3}
	\AddToShipoutPictureFG*{
		\put(\LenToUnit{#1\paperwidth},\LenToUnit{#2\paperheight}){\vtop{{\null}\makebox[0pt][c]{#3}}}
	}
}

\newcommand{\tquote}[1]{``#1''}

\begin{document}
	
\vspace*{0.2 cm}
\placetextbox{0.90}{0.97}{\small ULB-TH/21-07}

\title{Dark energy-dark matter interactions as a solution to the $S_8$ tension}

\author{Matteo Lucca}
\affiliation{Service de Physique Th\'{e}orique, Universit\'{e} Libre de Bruxelles, C.P. 225, B-1050 Brussels, Belgium}

\begin{abstract}
In this work we consider a scenario where the dark energy is a dynamical fluid whose energy density can be transferred to the dark matter via a coupling function proportional to the energy density of the dark energy. In particular, we investigate this model's ability to address the $S_8$ tension and find that against data from \textit{Planck}, BAO and Pantheon the model 1) can significantly reduce the significance of the tension, 2) does so without exacerbating nor introducing any other tension (such as the $H_0$ tension)  and 3) without worsening the fit to the considered data sets with respect to the $\Lambda$CDM model. We also test the model against data from weak lensing surveys such as KiDS and DES, and find that the model's ability to address the $S_8$ tension further improves, without a significant impact on any other parameter nor statistical measure. 
\end{abstract}

\maketitle

\section{Introduction}
Very little is known about the nature of dark energy (DE). In the standard cosmological model, the DE is nothing but a time-independent, spatially uniform, constant energy density (sometimes referred to as vacuum energy) contributing to approximately 70\% of the total energy budget of the universe today. This can be otherwise interpreted as a perfect simple fluid with equation of state (EOS) parameter $w_x=-1$. Yet, despite its simplicity, this parametrization allows for remarkably precise fits to the many available observables including, among many more, SNIa data, which provided the first evidence for an accelerated expansion of the universe \cite{Riess:1998cb, Perlmutter:1998np}, as well as the Cosmic Microwave Background (CMB) anisotropy power spectra and the late-time matter power spectra~\cite{Aghanim2018PlanckVI, Nadathur:2020kvq}. 

Nevertheless, in the effort to provide a deeper physical origin to the nature of the DE and often in connection to modified gravity, a number of attempts to go beyond this simple picture have been conducted in the literature (see e.g., \cite{Yoo:2012ug, Brax:2017idh} for in-depth reviews). For instance, one particularly well-motivated approach is to describe the DE as a dynamical fluid (also referred to as quintessence), with different possible time dependencies and interactions.

However, to this day no clear cosmological evidence for a deviation from the standard picture with a cosmological constant has been reported, and no statistically significant indication for a DE interaction has been found (see however \cite{Vagnozzi:2019kvw, Vagnozzi:2021quy} for possible alternative detection avenues). One of the main reasons for this is the strong synergy between Baryon Acoustic Oscillation (BAO) and SNIa data, whose combination strongly prefers $\Lambda$CDM over possible late-time modifications of the expansion history (see e.g., Fig.~3 of \cite{Poulin2018Implications} and related text for an overview). Additionally, it was pointed out in e.g.,~\cite{Bernal2016Trouble,Knox2019Hubble} that for a cosmological model to be able to successfully reconcile early- and late-time observations, in particular in relation to the Hubble parameter, it needs to modify the $H_0$ value \textit{and} the sound horizon at recombination time $r_s$, which is however only affected by the pre-recombination history of the universe, thus intrinsically disfavoring any late-time approach. Overall, a \textit{no-go theorem} seems to be emerging for late-time models predicting strong deviations from the $\Lambda$CDM scenario and in particular for those attempting to solve the Hubble tension (see e.g., Sec. 5 of \cite{DiValentino:2021izs} for a complete summary and e.g., \cite{Desmond:2019ygn} for possible exceptions).

For instance, one representative example of this behavior investigated at length in the literature is given by cosmological models allowing for a varying DE EOS parameter. A particularly detailed and data-driven analysis of this scenario can be found in e.g.,~\cite{Poulin2018Implications}, where the authors adopt a CPL parametrization of the EOS parameter, while a broader overview on other possible parametrizations is given in \cite{Yang:2021flj}, where it is shown that all models fail to alleviate the $H_0$ tension once BAO data is included. 

Nevertheless, as shown in e.g., \cite{Pourtsidou:2016ico, Camera:2017tws, Gomez-Valent:2018nib, Davari:2019tni, Benisty:2020kdt} as well as in this manuscript, there is still room for non-standard DE models to address another puzzling problem in modern cosmology, namely the discrepancy between the early-time inference and the low-redshift measurements of the amplitude of matter fluctuations, commonly parameterized with its value at a scale of 8~Mpc/$h$ referred to as~$\sigma_8$ (see e.g., \cite{DiValentino:2020vvd, Nunes:2021ipq} for concise overviews about the current status of this tension). In particular, here we will consider a model where the DE can interact with the dark matter (DM) via a coupling function proportional to the DE energy density. 

This type of interacting scenario has been extensively studied in the literature (see e.g., \cite{Gavela2009Dark, Gavela2010Dark} for the main original works and \cite{Wang2016Dark} for a recent review on DM-DE interactions in general), and its cosmological features diametrically differ depending on whether the energy density is flowing from the DE to the DM or \textit{vice versa}. Henceforth, we will refer to the two cases as iDEDM and iDMDE models, respectively. In a nutshell, in the former case, because of the additional energy injected from the DE into the DM over the cosmic history one has an overall suppression of the DM energy density with respect to the $\Lambda$CDM scenario. As a consequence, this results on the one hand in a decreased value of the Hubble parameter at late times, worsening the Hubble tension, and on the other hand in a delay of the radiation-matter equality from which follows a suppression of the matter power spectrum, which can alleviate the $\sigma_8$ tension. The opposite is true for the iDMDE case.

For these reasons, great attention has been dedicated to the iDMDE model in the context of the Hubble tension (see e.g., \cite{DiValentino2017Interacting, DiValentino2019Interacting, DiValentino2019Minimal}), although it has been explicitly shown in \cite{Lucca2020Shedding} that also this particular interacting DE scenario is unable to address the Hubble tension due to the aforementioned \textit{no-go theorem} that arises when BAO and SNIa data are accounted for. However, surprisingly, to our knowledge the iDEDM scenario has never been thoroughly investigated as a possible solution to the $\sigma_8$ tension (aside from a reference to a lowered $\sigma_8$ parameter in Tab. IV of \cite{DiValentino2019Minimal})\footnote{Attempts in this direction have been made in \cite{Kumar2019Dark, Kumar:2021eev} where the authors claim to be able to solve both the $H_0$ and the $\sigma_8$ tensions with negative values of the coupling constant, i.e., in the iDMDE model, and in \cite{An:2017crg} (see Model II there), where the authors claim to be able to solve the $\sigma_8$ tension despite actually increasing both the $\sigma_8$ (which would worsen the tension) and the $H_0$ values with respect to $\Lambda$CDM. In this sense, despite a seemingly similar mathematical setup at background level, their results appear to fundamentally contradict the aforementioned logic of mutual exclusion as a solution to either one tension or the other as well as the results discussed here and in e.g., \cite{DiValentino2017Interacting, DiValentino2019Interacting, DiValentino2019Minimal, Lucca2020Shedding} where that logic perfectly applies. One possible crucial difference might rely in the different sets of cosmological perturbation equations employed in~\cite{An:2017crg, Kumar2019Dark}, which seem to violate energy conservation, versus those used in the other references and explicitly derived in \cite{Gavela2009Dark, Gavela2010Dark}. Another possibly important variation could be in the choice of prior on the DE EOS parameter which is performed in \cite{DiValentino2017Interacting, DiValentino2019Interacting, DiValentino2019Minimal, Lucca2020Shedding} and here (and in \cite{An:2017crg}), but not mentioned at all in \cite{Kumar2019Dark, Kumar:2021eev} (the importance of this prior will be discussed below). In the case of \cite{An:2017crg}, one further source of concern is also the unreasonably high value of $H_0$ even if obtained with Planck 2015 data, which might point towards an inconsistency in their analysis. So, overall, because of some unexplained but obviously present underlying mathematical or numerical difference, we refrain from proposing a direct comparison to the results presented in \cite{Kumar2019Dark, Kumar:2021eev}, which were anyway developed in the context of the iDMDE model, while here we focus on the iDEDM model, and those presented in \cite{An:2017crg}, which were anyway obtained with a smaller and older sample of data sets compared to those included in this work.}. Therefore, we undertake this task here.

The paper is organized as follows. In Sec. \ref{sec: math} we briefly review the mathematical setup underlying the iDEDM model considered here. In Sec. \ref{sec: imp} we discuss the impact that the interacting model has on the matter power spectrum. In Sec. \ref{sec: meth} we introduce the numerical setup and cosmological probes used to evaluate the ability of the model to address the $S_8$ tension. In Sec. \ref{sec: res} we discuss the results on the basis of different data sets combinations. We conclude in Sec. \ref{sec: conc} with a summary and final remarks.

\section{Mathematical setup}\label{sec: math}
We begin by providing a brief overview of the mathematical setup underlying both the iDMDE and the iDEDM models. For sake of conciseness, we refer the interested reader to e.g.,~\cite{Lucca2020Shedding} and references therein (in particular \cite{Gavela2009Dark, Gavela2010Dark, DiValentino2017Interacting}) for more details.

At background level the effect of the DM-DE interaction is that the respective energy densities are not conserved independently, as in the standard $\Lambda$CDM case, but are coupled via an energy transfer function $Q$, reading
\begin{align}
	& \label{eq: rho_c} \dot{\rho}_{c}+3H\rho_{c}=Q\,, \\
	& \label{eq: rho_x} \dot{\rho}_x +3H\rho_x (1+w_x) =-Q\,,
\end{align}
where $\rho_i$ represents a fluid's energy density with indices $c$ and $x$ referring to the DM and the DE, respectively, and $H$ is the Hubble parameter. The energy transfer function $Q$ can be determined to be 
\begin{align}\label{eq: Q}
	Q=-\frac{\dot{\Lambda}}{8\pi G}
\end{align}
from Eq. \eqref{eq: rho_x}, recalling that $\rho_x=\Lambda/(8\pi G)$ and assuming that $\omega_x\simeq-1$. In the standard cosmological model, the value of $\Lambda$ is a constant, so that no energy transfer is present. However, within this work we will extend the $\Lambda$CDM model to allow for a time-dependent evolution of this quantity. In particular, as commonly done in the literature, we choose this dependence to be of the simple but arbitrary form
\begin{align}\label{eq: lambda}
	\Lambda=\Lambda_0(1+z)^\xi\,,
\end{align}
where $\Lambda_0=\Lambda(z_0)=3H_0^2\, \Omega_\Lambda$, $z$ is the redshift and $\xi$ is a free parameter of the model. Together with Eq.~\eqref{eq: Q}, Eq.~\eqref{eq: lambda} gives
\begin{align}\label{eq: Q 2}
	Q=\xi H\rho_x\,.
\end{align}
As clear form Eq. \eqref{eq: Q}, the sign of $\xi$ determines whether we are in the iDEDM (positive $\xi$) or in the iDMDE (negative $\xi$) scenario, i.e., whether the energy density is flowing from the DE to the DM or \textit{vice versa}, respectively.

A graphical representation of this setup is shown e.g., in Fig. 2 of \cite{Gavela2009Dark}. There, as mentioned above, the role of $\xi$ is that of scaling up or down the DM energy density in the past with respect to the $\Lambda$CDM model (depending on whether $\xi$ is negative or positive, and for fixed values of the DM and DE energy densities today), with an opposite effect on the DE energy density. One important consequence of this modified background evolution is then that the redshift of matter-radiation equality $z_{\rm eq}$ is advanced (with $\xi$ negative) or delayed (with $\xi$ positive), as can also be observed in the aforementioned figure.

The following perturbation equations read \cite{Gavela2009Dark, Gavela2010Dark, Honorez2010Coupled}
\begin{align}
	\label{eq: pert_1} \dot{\delta}_{c}= &-\theta_{c}-\frac{\dot{h}}{2}\left(1-\frac{\xi}{3} \frac{\rho_{x}}{\rho_{c}}\right) +\xi H \frac{\rho_{x}}{\rho_{c}}\left(\delta_{x}-\delta_{c}\right)\,, \\ 
	\label{eq: pert_3} \dot{\theta}_{c}=&-H\theta_{c}\,, 
\end{align}
\begin{align}
	\nonumber \dot{\delta}_{x}=&-(1+w_x)\left[\theta_{x}+\frac{\dot{h}}{2}\left(1+\frac{\xi}{3(1+w_x)}\right)\right]+ \\ & \hspace{0 cm} -3 H(1-w_x)\left[\delta_{x}+\frac{H \theta_{x}}{k^{2}}(3(1+w_x)+\xi)\right]\,, \\ 
	\label{eq: pert_2} \dot{\theta}_{x}=& 2 H \theta_{x}\left[1+\frac{\xi}{1+w_x} \left(1-\frac{\theta_{c}}{2\theta_{x}}\right)\right]+\frac{k^{2}}{1+w_x} \delta_{x}\,,
\end{align}
with initial conditions for the DE perturbations given by~\cite{Salvatelli2013New}
\begin{align}
	\delta_{x}^{in}(x) =(1+w_x-2 \xi)C  \hspace{0.4 cm}\text{and}\hspace{0.4 cm} \theta_{x}^{in} =k^2\tau C\,,
\end{align}
where
\begin{equation}
	C=-\frac{1+w_x+\xi / 3}{12 w_x^{2}-2 w_x-3 w_x \,\xi+7 \xi-14}\frac{2 \delta_{\gamma}^{in}}{1+w_{\gamma}}\,.
\end{equation}
Here, $\delta=\delta \rho/\rho$ with $\delta \rho$ as the first-order perturbation of the energy density $\rho$, $\theta$ is the fluid's divergence velocity, $h$ is the trace part of the metric perturbation (not to be confused with the dimensionless Hubble parameter in Eq.~\eqref{eq: k_eq}), and $k$ represents the Fourier modes. As commonly done in the literature \cite{Gavela2010Dark, DiValentino2017Interacting, DiValentino2019Interacting, DiValentino2019Minimal, Lucca2020Shedding}, here we have neglected the center of mass velocity for the total fluid $v_T$ and fixed the DE sound speed to unity, i.e., $c_{s,x}^2=1$, while for the DE adiabatic sound speed we have $c_{a,x}^2=w$ (see e.g., Sec.~2.3 of \cite{Valiviita2008Large} for more details as well as Appendix C of \cite{Lucca:2021eqy} for a careful derivation). 

In the analysis of these equations, it has been realized in \cite{Gavela2009Dark, Gavela2010Dark} that, even if the value of $w_x$ is kept fixed, its choice has to be made with care. The reason for this is a combination of factors. Firstly, as clear from Eqs. (10)-(11) of \cite{Lucca2020Shedding}, if $w_x$ was exactly equal to $-1$ (as in the standard scenario), the perturbation equations would diverge (effect also known as gravitational instability). Furthermore, following Secs.~2.2 and 3.4 of \cite{Gavela2009Dark}, it turns out that early-time instabilities can be avoided only if $\xi$ and $w_x+1$ are of opposite sign, i.e., if the so-called doom factor introduced in the reference is negative. Therefore, when fixing the DE EOS parameter in the iDMDE and the iDEDM models one commonly employs the values of $-0.999$ and $-1.001$, respectively. These values are chosen because they are close enough to the standard value to recover $\Lambda$CDM when $\xi=0$, but avoid both gravitational and early-time instabilities at the same time.

\section{Impact on the observables}\label{sec: imp}
Now that the main equations have been outlined, we can turn our attention to the impact that DE-DM interactions have on the observables, focusing in particular on the matter power spectrum since a detailed interpretation of the physical effects related to the CMB temperature anisotropy power spectrum can already be found in e.g.,~\cite{Murgia2016Constraints}. For the numerical evaluation of the many quantities involved we will employ the modified CLASS~\cite{Lesgourgues2011Cosmic, Blas2011Cosmic} version made publicly available in \cite{Lucca2020Shedding}.

\begin{figure}[t]
	\centering
	\includegraphics[width=\columnwidth]{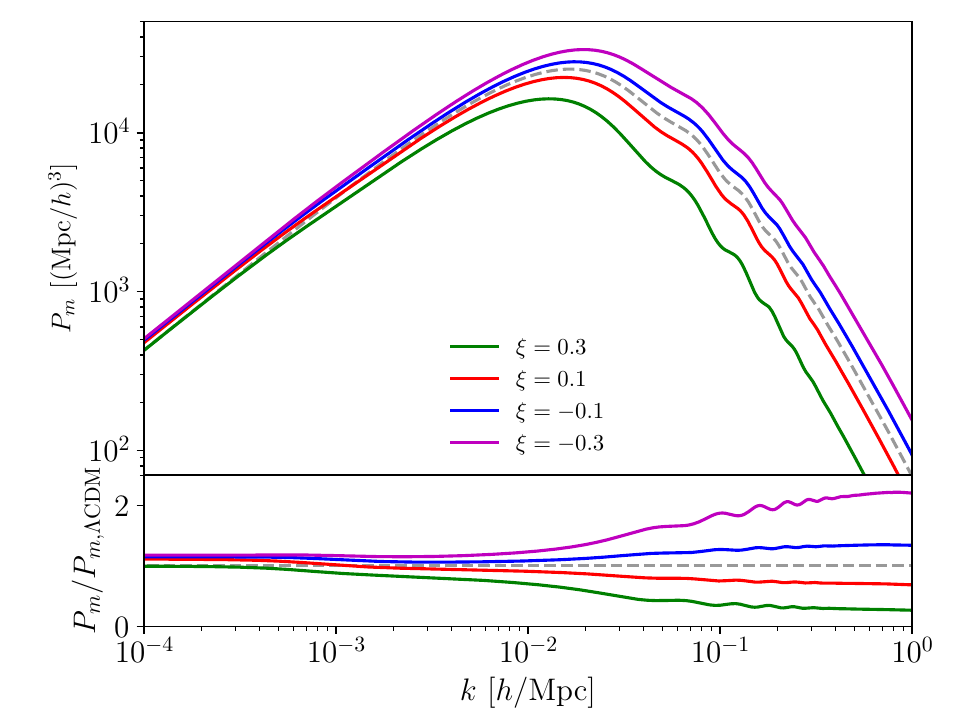}
	\caption{Effect of varying $\xi$ on the matter power spectrum (with the standard $\Lambda$CDM parameters fixed to the Planck+BAO best fits \cite{Aghanim2018PlanckVI}) for both the iDEDM ($\xi$ positive) and iDMDE ($\xi$ negative) models. The lower panel shows the same quantities as the upper panel, but normalized to the $\Lambda$CDM equivalent ($\xi=0$), which is shown as a gray dashed line for reference. The dashed green line is shown to highlight the degeneracy between $\xi$ and $\omega_c$ discussed in the text.}
	\label{fig: evolution_Pk}
\end{figure}

A graphical representation of the effects that varying $\xi$ has on the matter power spectrum can be seen in Fig.~\ref{fig: evolution_Pk} (solid lines). There, one immediately recognizes two very strong effects. The first is an horizontal shift of the peak scale and the second is an overall suppression or enhancement of the amplitude of the spectrum (small at scales below the peak and much larger at higher scales) depending on whether $\xi$ is positive or negative.

The first effect can be understood following the evolution of the scale of equality $k_{\rm eq}$, defined as
\begin{align}\label{eq: k_eq}
	\nonumber k_{\rm eq} & = a_{\rm eq}H_{\rm eq} \simeq \frac{1}{1+z_{\rm eq}}\frac{H_0}{h}\sqrt{2\omega_m(1+z_{\rm eq})^3} \\
	& \propto \sqrt{\omega_m(1+z_{\rm eq})}\,,
\end{align}
where we neglected the DE contribution, and made use of the knowledge that $k_{\rm eq}$ corresponds to the scale of the peak of the power spectrum (see e.g., Sec. 3.2 of \cite{Lesgourgues:2013qba}). Here, $a$ is the scale factor (labeled $a_{\rm eq}$ when taken at matter-radiation equality), $H$ represents the Hubble parameter (labeled $H_0$ when taken today and $H_{\rm eq}$ when taken at matter-radiation equality) and $h$ its dimensionless form, such that $H_0/h = 100$ km/(s Mpc), while $\omega_{m}=\Omega_{m}h^2$ is the dimensionless matter energy density (the same notation will be used in the following text also in relation to the DM, indexed as $c$). Then, for $\xi$ positive (negative) we have that both $\omega_m$ and $z_{\rm eq}$ decrease (increase), as clear from the discussion above and e.g., Fig. 2 of \cite{Gavela2009Dark}. As a results, $k_{\rm eq}$ decreases (increases) and therefore the position of the peak shifts to the left (right).

On the other hand, in order to capture the physics behind the vertical shifts of the spectrum we need to consider mainly three competing effects at play at the same time. First of all, for $\xi$ positive or negative, the larger the value of $\xi$ is, the more the scale of DE-matter equality will shift to the left (smaller scales) or to the right (larger scales)\footnote{Recall that, assuming $\xi$ to be positive (negative), we have both that the DE energy density increases (decreases) and that the DM energy density decreases (increases).}, and the larger the enhancement or suppression of the spectrum below \text{$k_\Lambda\simeq5\times10^{-4}$ $h$/Mpc} will be (see e.g., Fig.~14 of \cite{Lesgourgues:2013qba} for a graphical depiction of this effect). At the same time, however, for $\xi$ positive or negative, the larger the value of $\xi$ is, the smaller or larger the value of the conformal time today $\eta_0$ is, and correspondingly also the overall amplitude of the power spectrum which is directly proportional to $\eta_0$ \cite{Lesgourgues:2013qba}. Finally, for scales above the baryon drag scale (corresponding to the baryon drag time) the matter power spectrum also assumes an additional dependence of the form ${k_{\rm eq}^2(1-\omega_b/\omega_{m})^2}$ \cite{Lesgourgues:2013qba}. In this way, for $\xi$ positive or negative, the larger the value of $\xi$ is, the larger the suppression or enhancement of the power spectrum at large scales.

The combination of these effects is clearly visible in Fig.~\ref{fig: evolution_Pk} (solid lines). Indeed, for positive values of $\xi$ at very small scales one observes the enhancement due to the first effect explained above (modified time of DE-matter equality) dominating the suppression due to the second effect  (modified overall amplitude) up until $\xi=0.3$, while above this value and for negative values of $\xi$ the second effect always dominates. Then, at $k$ values above $k_\Lambda$ one has the expected suppression or enhancement of the power spectrum depending on whether $\xi$ is positive or negative, behavior that becomes more pronounced at scales roughly above~$k_{\rm eq}$.

All of these effects are, however, almost completely degenerate with what is expected from an increased DM energy density, as also discussed and illustrated in e.g.,~\cite{Murgia2016Constraints} for the CMB anisotropy power spectra (see Sec. 2.2 and Fig. 1 therein). Indeed, it is enough to increase the values of $\omega_c$ today to compensate for the suppressed background evolution (in particular at early times) and restore the standard value of~$z_{eq}$, which is the quantity most strongly driving the impact on the observables due to the presence of the DE-DM interactions. To graphically show this degeneracy in the context of the matter power spectrum, we plot in Fig. \ref{fig: evolution_Pk} a representative example (green dashed line) with $\xi=0.3$ and $\omega_c=0.15$ (otherwise fixed to $\omega_{c}\simeq0.12$) and indeed the curve nicely overlaps with the $\Lambda$CDM prediction (gray dashed line).

Therefore, the mechanism through which the iDEDM model could potentially achieve a suppression of the amplitude of matter fluctuations is to reduce the amount of DM in the past by means of the DE-DM interactions and correct for it by increasing the DM energy density today but only to a point where $z_{eq}$ and $k_{eq}$ are still reduced with respect to the $\Lambda$CDM model.

\section{Method and cosmological probes}\label{sec: meth}
As explained in the previous section, for positive values of $\xi$, i.e., within the iDEDM model, one observes an overall suppression of the matter power spectrum at all scales (up to a negligible enhancement at $k<k_\Lambda$ for $\xi<0.3$), which could in principle help to alleviate the $\sigma_8$ tension. To quantitatively evaluate this expectation we use the aforementioned modified version of CLASS in combination with the parameter inference code MontePython \cite{Audren2013Conservative, Brinckmann2018MontePython} to perform a series of Markov chain Monte Carlo (MCMC) scans over the free parameters of the model. In particular, we consider a 6+1 extension of the standard $\Lambda$CDM model with
\begin{align}\label{eq: params set e1}
	\{H_0,\, \omega_b,\, \omega_{c},\, n_s,\, \ln(10^{10}A_s),\, \tau_{\rm reio} \} + \xi\,,
\end{align}
imposing the condition $\xi>0$ and hence ${w_x=-1.001}$, as explained above. Here, $A_s$ and $n_s$ represent the amplitude of the primordial power spectrum and its scalar spectral index, respectively, while  $\tau_{\rm reio}$ is the reionization optical depth.

In order to constrain this set of parameters we consider the same cosmological data sets as in \cite{Lucca2020Shedding}, namely the CMB mission \textit{Planck} 2018 \cite{Aghanim2018PlanckVI} (including the temperature, polarization and lensing likelihoods, henceforth only referred to as \textit{Planck}), the BAO data gathered by the 6dF Galaxy Survey (6dFGS) via the measurement of $D_V/r_{\rm drag}$ at $z=0.106$ \cite{Beutler2011Galaxy}, by the Sloan Digital Sky Survey (SDSS) from the MGS galaxy sample at $z=0.15$ \cite{Ross2014Clustering}, and by the Baryon Oscillation Spectroscopic Survey (BOSS) from the CMASS and LOWZ galaxy samples of SDSS-III DR12 at $z=0.2-0.75$ \cite{Alam2016Clustering}, as well as by the SNIa Pantheon catalog \cite{Scolnic2017Complete}. However, here we refrain from including any prior on $H_0$ coming from late-time measurements such as the one reported by the SH0ES collaboration \cite{Riess2019Large}, following the arguments recently pointed out in e.g., \cite{Benevento:2020fev, Camarena:2021jlr, Efstathiou:2021ocp}. Henceforth, we will refer to the combination of \textit{Planck}, BAO and Pantheon as \textit{baseline}. Clustering constraints such as redshift-space distortions are also not included in the current analysis.
\begin{table*}[t]
	\centering
	\begin{tabular}{|c|c|c|c|c|}
		\hline\rule{0pt}{3.0ex} 
		\multirow{2}{*}{Parameter} & \multicolumn{2}{|c|}{\textit{baseline}} & \multicolumn{2}{|c|}{\textit{baseline}+KV450+DES} \\[0.1 cm]
		\cline{2-5}
		\rule{0pt}{3.0ex}
		& $\Lambda$CDM & iDEDM & $\Lambda$CDM & iDEDM  \\[0.1 cm]
		\hline
		\rule{0pt}{3.0ex}
		$H_0 \, [\text{km/(s Mpc)}]$ & $68.09\pm0.43$ & $67.63_{-0.51}^{+0.62}$ & $68.65_{-0.39}^{+0.40}$ & $68.29_{-0.46}^{+0.52}$ \\[0.1 cm]
		$\omega_c$           & $0.11954_{-0.00095}^{+0.00094}$ & $0.1258_{-0.0061}^{+0.0027}$ & $0.11830_{-0.00088}^{+0.00081}$ & $0.1230_{-0.0047}^{+0.0020}$ \\[0.1 cm]
		$\Omega_{m}$                 & $0.3062_{-0.0057}^{+0.0055}$ & $0.3243_{-0.019}^{+0.0097}$ & $0.2988_{-0.0052}^{+0.0049}$ & $0.3122_{-0.015}^{+0.0078}$ \\[0.1 cm]
		$\xi$                        & - &  $<0.16$ & - &  $<0.12$ \\[0.1 cm]
		$\sigma_8$                   & $0.8224_{-0.0061}^{+0.0059}$ & $0.783_{-0.018}^{+0.039}$ & $0.8156_{-0.0057}^{+0.0055}$ & $0.787_{-0.013}^{+0.030}$ \\[0.1 cm]
		$S_8=\sigma_8(\Omega_m/0.3)^{0.5}$ & $0.831_{-0.010}^{+0.011}$ & $0.813_{-0.014}^{+0.019}$ & $0.8139_{-0.0093}^{+0.0092}$ & $0.802_{-0.012}^{+0.015}$ \\[0.1 cm]
		\hline
		\rule{0pt}{3.0ex}
		$\Delta\chi^2$               & - & $-1.6$ & - &  $-0.4$ \\[0.1 cm]
		$\sigma$             		 & - &  $1.3$ & - &  $0.6$ \\[0.1 cm]
		$\ln B_{\rm iDEDM,\Lambda}$          & - &  $-2.2$ & - &  $-2.4$ \\[0.1 cm]
		\hline
	\end{tabular}
	\caption{Mean and 68\% C.L. of the parameters most significantly affected by the iDEDM model (the upper bounds are given at the 95\% C.L.) for different models and data set combinations, together with the corresponding $\Delta\chi^2$ value, the statistical significance $\sigma$ and the Bayes ratio $\ln B_{\rm iDEDM,\Lambda}$ (evaluated with respect to the $\Lambda$CDM model).\vspace*{3mm}}	
	\label{tab: MCMC_res}
\end{table*}

Furthermore, since when considering the \textit{baseline} data sets the iDEDM model is able to restore the concordance between early-time inference and late-time measurements of the $\sigma_8$ parameter (as shown in Sec.~\ref{sec: res}), it becomes in principle statistically possible to additionally include weak-lensing data gathered by the combination of the Kilo-Degree Survey and the VISTA Kilo-Degree Infrared Galaxy Survey (henceforth referred to as KV450) \cite{Hildebrandt:2018yau} as well as by the Dark Energy Survey (DES)~\cite{Abbott:2017wau, Troxel:2017xyo}. To account for the former we employ the combined likelihood distributed with the public version of MontePython and based on \cite{Hildebrandt:2018yau}, for which we only rely on the linear scales since a dedicated treatment of the non-linear scales for the model considered in this work is currently not available in the literature. In absence (to our knowledge) of a publicly available DES likelihood, we parameterize the impact of this probe as a Gaussian prior on the parameter ${S_8=\sigma_8(\Omega_m/0.3)^{0.5}}$, which is possible since the values taken by $\Omega_m$ in Fig. \ref{fig: MCMC_res} always fall within the posterior distribution shown in e.g., Fig. 7 of \cite{Troxel:2017xyo}. Following \cite{Troxel:2017xyo}\footnote{Note that also other evaluations of the DES data are present in the literature when combined with KV450 data \cite{Joudaki:2019pmv, Asgari:2019fkq}, with different degrees of physical and mathematical refinements. However, as the results presented there can vary quite significantly in particular with respect to the values of $\Omega_{m}$, here we base ourselves only on the original results presented in \cite{Troxel:2017xyo}. A qualitative discussion accounting for the findings of \cite{Joudaki:2019pmv, Asgari:2019fkq} will follow later in the text.}, one has that $S_8=0.782\pm0.027$.

Note that, in principle, one could have also considered other (combinations of) measurements and surveys, such as the Subaru Hyper Suprime-Cam (HSC) survey \cite{Hikage:2018qbn} or KiDS-1000 \cite{Heymans:2020gsg}, but in these cases (to our knowledge) public MontePython-compatible likelihoods are missing and the posterior distributions displayed in the release papers do not follow the same Gaussian behavior as in the case of DES, making a simple extraction of a likelihood very complicated. We also refrain from including information from e.g., BOSS because of the currently very large error bars on $S_8$ and~$\Omega_{m}$, which do not significantly affect the KV450 estimate when combined~\cite{Troster:2019ean}.

Finally, we also evaluate the statistical significance of the iDEDM model with respect to the $\Lambda$CDM model using a standard $\Delta \chi^2$ comparison of the best-fit models (which relies only on the goodness of the fit to the different data sets), the significance $\sigma$ (which also accounts for the various degrees of freedom) and the Bayes ratio (which also includes information on the priors of the models) expressed as $\ln B_{\rm iDEDM,\Lambda}$ and computed using the numerical code \texttt{MCEvidence} \cite{Heavens:2017afc} (see e.g., Sec. V of \cite{Gomez-Valent:2020mqn} -- which we follow in notation and method -- for a concise but very instructive discussion around this quantity).

\section{Results}\label{sec: res}
The results obtained with this procedure are summarized in the left column of Tab. \ref{tab: MCMC_res}, where only the parameters most impacted by the presence of the DE interactions are displayed. For the \textit{baseline} data sets, from the table it is first of all clear that the iDEDM model can significantly reduce the $\sigma_8$ value (which we find to be $\sigma_8=0.8224_{-0.0061}^{+0.0059}$ for $\Lambda$CDM with the same data sets) without decreasing the $H_0$ value significantly\footnote{As a remark, note also that the eventual inclusion of the DE EOS $w_x$ as a free parameter is not expected to impact this conclusion significantly. Indeed, even in this case, despite not employing the very same data sets considered here, both \cite{DiValentino2019Minimal} (see Tab. IV there where however BAO and SNIa are not shown combined) and \cite{Martinelli:2019dau} (where Planck 2015 data was employed) find very similar upper bounds on $\xi$ as the one shown in Tab. \ref{tab: MCMC_res}.}. This is already remarkable given the inability of the iDMDE model to achieve a similar result when attempting to address the $H_0$ tension (see e.g., Tab. of II of \cite{DiValentino2019Minimal}). 

However, this comes at the cost of an increased value of $\omega_c$ and hence of $\Omega_{m}$ (the value of $\omega_b$ remains unaltered), so that the value of $S_8=\sigma_8(\Omega_m/0.3)^{0.5}$, which best captures the  degeneracy between $\sigma_8$ and $\Omega_{m}$, is not as significantly reduced as the $\sigma_8$ value alone is. As explained in Sec. \ref{sec: imp}, the increase in $\omega_c$ is needed in order to compensate for the suppression introduced by the DE-DM interactions of the DM density at the time of matter-radiation equality. However, as hinted to in Sec.~\ref{sec: imp}, the data still allows for an overall reduction of the DM energy density at around $z_{eq}$, so that a decrease of the amplitude of the matter power spectrum is possible in the iDEDM model. Indeed, despite the increased value of $\omega_{c}$, the value of $S_8$ obtained in the context of the iDEDM model is still considerably lower than the value  one would obtain in the $\Lambda$CDM model for the same data set combination, which amounts to $S_8=0.831_{-0.010}^{+0.011}$ (see Tab. \ref{tab: MCMC_res}). 

Therefore, with respect to the value of $S_8$ reported for instance by~\cite{Hildebrandt:2018yau} for KV450, i.e., $S_8=0.737_{-0.036}^{+0.040}$, the tension between early and late-time probes reduces from 2.4$\sigma$ in the $\Lambda$CDM model to 1.8$\sigma$ in the iDEDM model. Similarly, compared to the value reported by \cite{Troxel:2017xyo} for DES, i.e., $S_8=0.782\pm0.027$, one has a reduction of the discrepancy from 1.7$\sigma$ to 1.0$\sigma$. Remarkably, even very discordant estimates such as the ones reported in~\cite{Ivanov:2019pdj} and \cite{DAmico:2019fhj} for alternative analysis of the BOSS data, i.e., $S_8=0.703\pm 0.045$ and $S_8=0.704\pm 0.051$, respectively, are brought below a 2.3$\sigma$ significance (while being at about $2.8\sigma$ in the $\Lambda$CDM model). 

\begin{figure}
	\centering
	\includegraphics[width=\columnwidth]{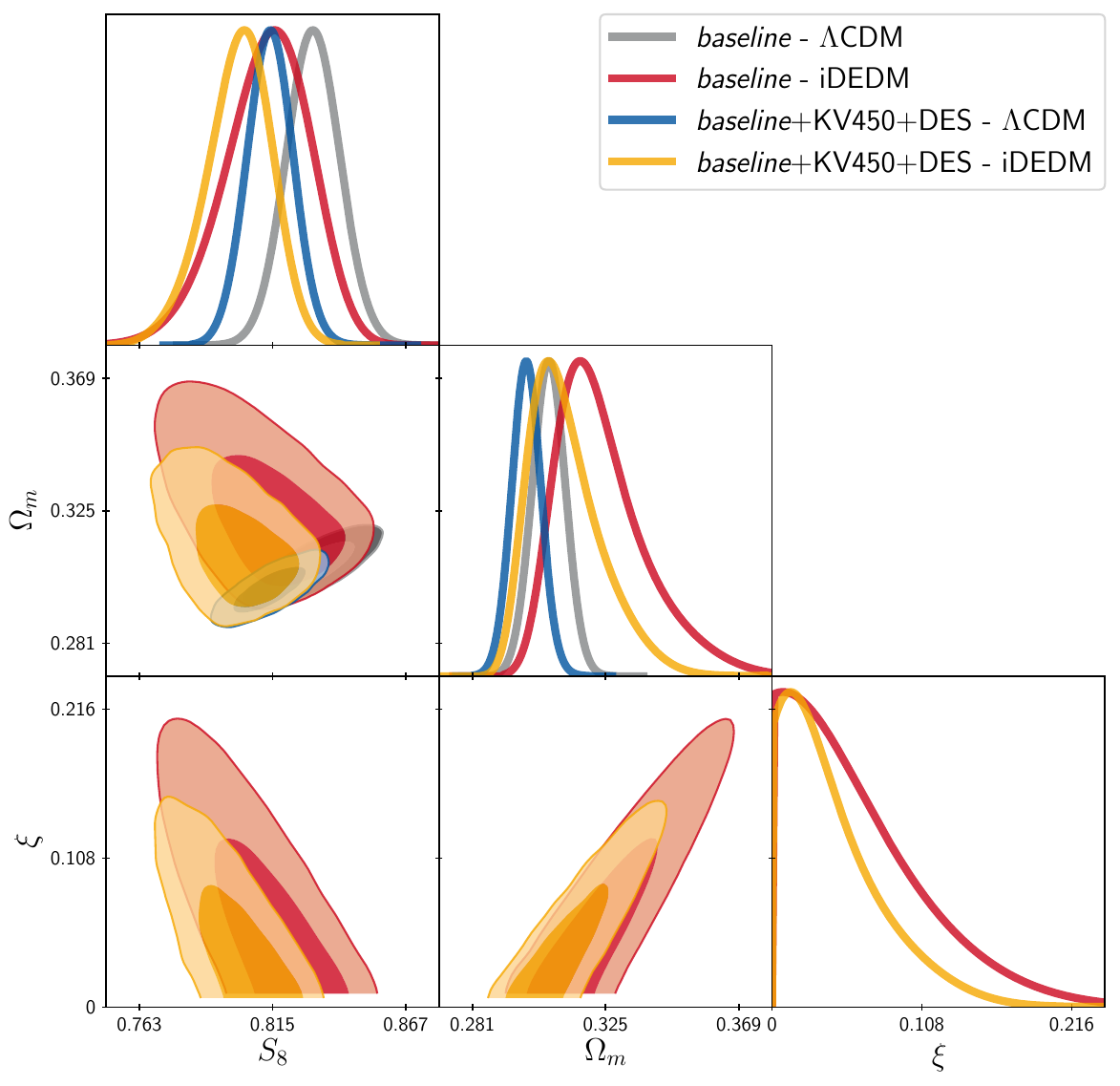}
	\caption{Two-dimensional contours (68\% and 95\% C.L.) of the $S_8-\Omega_{m}-\xi$ plane for different models ($\Lambda$CDM and iDEDM) and data sets combinations (\textit{baseline} and \textit{baseline}+ KV450+DES).}
	\label{fig: MCMC_res}
\end{figure}

The same qualitative conclusions can also be drawn from the comparison between the gray and red contours in Fig. \ref{fig: MCMC_res}, where the posterior distributions  for the $S_8-\Omega_{m}-\xi$ plane are shown. Indeed, there one observes the increased value of $\Omega_{m}$ in the iDEDM model (red) with respect to $\Lambda$CDM (gray) and the strong degeneracies this parameter (or better, $\omega_c$) shares with $\xi$ and $S_8$: the higher the value of $\xi$, the lower the value of $S_8$ and the higher the value of $\Omega_{m}$, as expected.

Furthermore, for the \textit{baseline} data sets the iDEDM model reduces the $\chi^2$ by about 1.6 with respect to the $\Lambda$CDM model, which corresponds to a negligible 1.3$\sigma$ preference for the former model over the latter, while the Bayesian evidence is just as negligible (-2.2) but in favor of $\Lambda$CDM.

Overall, the iDEDM model appears to be very effective in lowering the significance of the tension between the inference of the $S_8$ parameter from the \textit{baseline} data sets and its many late-time measurements. This is even more appealing since it happens without deteriorating neither the fit to the early-time probes nor the Hubble tension, and it does not require any late-time information. 

However, there is one important aspect of the analysis that needs to be investigated more in depth. Indeed, although the values of $\Omega_{m}$ obtained in the iDEDM model for the \textit{baseline} data sets are well within the distributions shown in \cite{Hildebrandt:2018yau} and \cite{Troxel:2017xyo} for KV450 and DES, respectively, these data sets prefer the lowest end of the distributions of $\Omega_{m}$ rather than the higher values which allow the iDEDM model to address the $S_8$ tension so successfully\footnote{Notably, although in the attempt to address another tension, the $H_0$ tension, this issue also appears in the context of Early Dark Energy \cite{Poulin2018Early, Smith2019Oscillating}, as discussed in depth in e.g., \cite{Hill2020Early}. A comparatively detailed investigation will not be conducted within this manuscript, but will serve as guideline for future efforts.}. This trend is also confirmed by recent attempts to combine the two data sets \cite{Joudaki:2019pmv, Asgari:2019fkq}. 

Therefore, in order to evaluate the extent of the impact of this preference in the value of $\Omega_{m}$ we additionally include data gathered by KV450 and DES~\cite{Hildebrandt:2018yau, Abbott:2017wau, Troxel:2017xyo} as explained in Sec. \ref{sec: meth}. The results for the \textit{baseline}+KV450+DES case are shown in the right columns of Tab. \ref{tab: MCMC_res} and in Fig. \ref{fig: MCMC_res} (blue and orange contours). As one can see there, the values of both $S_8$ and $\Omega_{m}$ are lowered, as expected, and so is the upper bound on $\xi$. As a consequence, the tensions with KV450 and DES decrease to 1.6$\sigma$ and 0.7$\sigma$, respectively, without any appreciable change in any of the other cosmological parameters nor statistical measure. 

This is an important result considering that, as hinted to before, in the $\Lambda$CDM model one has intrinsically lower values of $\omega_c$ as compared to the iDEDM model, so that the stronger incompatibility in $S_8$ could have been balanced by a better agreement in the values of $\Omega_{m}$ and the goodness of the fit to the data as a whole might have suffered from the higher values of $\Omega_{m}$ in the iDEDM model. As a remark, however, although this is true with the inclusion of the KV450+DES data sets, this behavior would likely be exacerbated by the inclusion of additional probes such as the Subaru HSC survey (see the distribution in e.g., Fig. 4 of \cite{Hikage:2018qbn}) or the reanalysis of the KV450+DES data as presented in~\cite{Asgari:2019fkq} (see e.g., Fig. 5 there). Nevertheless, we refrain from speculating too freely on this aspect, leaving a more in-depth investigation for future work, possibly with the inclusion of official likelihoods.

\section{Conclusions}\label{sec: conc}
In this manuscript we have considered a cosmological model allowing for DE-DM interactions (with coupling function proportional to the DE energy density), where the energy density flows from the DE to the DM (i.e., the coupling parameter is assumed to be positive). This scenario has been largely studied in the literature in relation to the Hubble tension, but never as a solution to the $\sigma_8$ tension. Nevertheless, as we explain in the text, this interacting model is in principle able to delay the time of matter-radiation equality by suppressing the amount of DM energy density in the past with respect to $\Lambda$CDM, and thereby to lower the value of the corresponding scale which ultimately determines the position of the peak and the amplitude of the matter power spectrum. Therefore, via this mechanism this particular model can suppress the amplitude of matter fluctuations and hence decrease the $\sigma_8$ value.

We have therefore quantitatively tested this model's ability to address the aforementioned tension considering data from \textit{Planck}, BAO and SNIa. We find that the interacting DE-DM scenario considered here is indeed capable of restoring to a large extent the concordance between early-time inference and late-time measurements of the $S_8$ parameter (dropping below the $2.5\sigma$ level even for very discordant estimates). Furthermore, importantly, this happens without worsening the Hubble tension in the process nor the goodness of the fit to the data. We further investigated the impact of adding data from weak lensing surveys such as KV450 and DES, and found that the $S_8$ values further lowers without significantly affecting any other parameter nor statistical measure.

This successful behavior could then possibly set the seed for an even more inclusive cosmological model able to address both the $H_0$ and $S_8$ tensions at the same time with a combination of early- and late-time extensions of the $\Lambda$CDM model.

\section*{Acknowledgements}
We thank Deanna C. Hooper and Sunny Vagnozzi for the many very insightful discussions. This work was supported by an F.R.S.-FNRS fellowship, by the \tquote{Probing dark  matter with neutrinos} ULB-ARC convention and by the IISN convention 4.4503.15. Computational resources have been provided by the Consortium des Équipements de Calcul Intensif (CÉCI), funded by the Fonds de la Recherche Scientifique de Belgique (F.R.S.-FNRS) under Grant No. 2.5020.11 and by the Walloon Region.

\newpage

\bibliography{bibliography}{}

\end{document}